\documentstyle[11pt,newpasp,twoside,epsf,psfig]{article}
\def\edcomment#1{\iffalse\marginpar{\raggedright\sl#1\/}\else\relax\fi}
\reversemarginpar

\begin{document}
\title{Determining the Properties of Galaxy 2237+0305 Using Gravitational Lensing}
 \author{Cathryn Trott, Rachel Webster}
\affil{School of Physics, The University of Melbourne, Victoria 3010 AUSTRALIA}

\begin{abstract}
We have studied the mass distribution in the lensing galaxy 2237+0305 using constraints from both gravitational lensing and photometric and spectroscopic observations. We find that with sufficient dynamical information we can constrain the mass-to-light ratio of the luminous components and determine the dynamical contribution of the disk. In addition, future observations should allow us to place constraints on the shape of the dark matter halo and its inner slope.
\end{abstract}
\vspace{-4mm}
\section{Introduction}
Gravitational lensing has become an important tool in recent years for studying many aspects of the universe. As it probes total mass, independent of the light distribution, it opens new windows onto studies of large scale structures where dark matter has dynamical importance.

With new galaxy lenses being discovered each year, the statistical use of systems to study dark matter halo shapes and orientations is fast becoming possible. Most lensing systems lie at high redshift as this is where the lensing cross-section is optimal, however for detailed studies of the lensing galaxies themselves, lower redshift systems are better laboratories.

2237+0305 was discovered as a low redshift quadruple lens by Huchra et al. (1985). At $z$=0.039 it is a visibly extended early-type spiral galaxy with multiple luminous stellar components. In addition, its proximity to us places its four lensed quasar images close to the centre of the galaxy at $\sim$700pc (most lens images sit $\sim$5kpc from the deflector's centre) making the probed region of the galaxy more compact. These features of the 2237+0305 system make it an ideal laboratory for studying the inner structure of the dark matter halo and properties of the lensing galaxy.

Most galaxy lenses are early-type and studies of these systems can also provide important constraints on the dark matter (for example Treu, this meeting). Studies with spiral lenses are less frequent due to the fewer number known (elliptical galaxies present a larger lensing cross-section probability) but can provide information about the luminous components of the galaxy (for example the system B1600+434, Maller et al. 2000).

The importance of the stellar disk to the dynamical support of the galaxy is a controversial question in modern galactic astronomy. Sackett (1997) defined a maximal disk as one which provides 75-95 per cent of the galaxy's dynamical support. Whether a disk is maximal or sub-maximal has an impact on our understanding of disk dynamics and is therefore an important quantity to determine. Similarly, the mass-to-light ratios in the stellar components of a late-type galaxy are not well determined and their values can place constraints on star formation models. Traditional studies of spiral galaxies have used light profiles of the luminous components and spectroscopic dynamical information to determine these properties of the galaxy. In a complex system with multiple mass components, a degeneracy exists between the mass-to-light ratio of the major stellar component and the mass scaling of the dark matter halo. The additional information provided by lensing can break this degeneracy.
\vspace{-4mm}
\section{Galaxy Modelling}
2237+0305 has been studied extensively in the twenty years since its discovery (for example Yee 1988, Foltz et al. 1992, Schmidt 1996). Many simplified models of the galaxy's mass distribution have been published which focus on different aspects of the system. We aim to model the major mass components with physically reasonable profiles and use these models to constrain the mass-to-light ratios, disk maximality and dark matter halo inner slope.

Schmidt (1996) was the first to investigate the effect of the galaxy's bar on the lensed image positions. He found the bar provides a significant torque to their positions and is thus a critical component to any lens model. In addition to the bar, we model an inclined stellar disk, an elliptical bulge at the same position angle, and a dark matter halo. The disk is modelled as a thin structure with an exponential surface density profile, the bulge with a de Vaucouleurs profile and the bar as a Ferrers ellipse, in line with previous works. The dark matter halo we model with both a flattened NFW profile (Navarro, Frenk \& White 1996) and as a softened isothermal sphere (SIS). The parameters characteristic to these models have been taken from previous observational studies and are presented in Table 1.
\begin{table}
\begin{center}
\begin{tabular}{ll|ll}
{\bf DISK} & M/L = free & {\bf BAR} & M/L = free\\
 & r$_d$ = 11.3$\pm$1.2$\arcsec$\footnotemark[1]&  & e = 0.64\footnotemark[1]\\
 & i = 60$\deg$\footnotemark[4] &  & PA = 39$\deg$\footnotemark[2]\\
 & PA = 77$\deg$\footnotemark[2] & & \\\hline
{\bf BULGE} & M/L = free &  {\bf DM HALO} & M/L = free\\
 & r$_b$ = 4.1$\pm$0.4$\arcsec$\footnotemark[1] & SIS & r$_c$ = free\\
 & e = 0.31\footnotemark[3] & eNFW & r$_h$ = free\\
 & PA = 77$\deg$\footnotemark[2] & & e = free\\
\end{tabular}\caption{Fixed and free parameters and their source for the mass models of the four major components. The label `free' refers to free parameters in the model and $r_c$ and $r_h$ denote the core radius and scale length respectively. $^1$Schmidt 1996, $^2$Yee 1988, $^3$Racine 1991, $^4$Irwin et al. 1989.}
\end{center}
\end{table}
The source position remains unknown and increases the number of free parameters by two. In addition to the constraints from observations, the four image positions are well determined by $HST$ (positions of Blanton, Turner \& Wambsganss 1998 used in this study). Barnes et al. (1999) observed the neutral hydrogen 21cm emission line from the galaxy with the $VLA$ and this low resolution data has allowed us to obtain two independent points in the outer rotation curve.

With the known image positions and rotation curve information, there are ten constraints on the model. For the softened isothermal sphere, there are seven free parameters and for the flattened NFW there are eight. The problem is therefore over-constrained.
\vspace{-4mm}
\section{Inversion Method}

The unknown parameters are varied to fit both the image positions and rotation curve points using a basic minimisation procedure and the $\chi^2$ parameter. The disk and bulge scale lengths are also allowed to move within their 1$\sigma$ uncertainties in order to explore the parameter space adequately. The bending angles for each mass component are computed and their relative contributions (mass-to-light ratios) varied to fit the constraints. Solutions with unphysically large or small values for the mass-to-light ratios were discounted. Since most components contribute shear as well as convergence to the lensing configuration, degeneracies exist between some parameters.
\vspace{-4mm}
\section{Results}

Statistically acceptable solutions were found for both dark matter halo density profiles. The degree of ellipticity required for the flattened NFW profile was very small and consistent with zero. Both solutions provided similar convergence within the image regions and therefore allowed the same combination of the luminous mass components.

For 3 (2) degrees of freedom, an acceptable $\chi^2$ value at 1$\sigma$ confidence is 3.6 (2.4). Table 2 shows the best-fitting parameters for the models.
\begin{table}
\begin{center}
\begin{tabular}{l|ll}
 & SIS & eNFW\\ \hline
M/L$_{disk,I}$ & 0.74$\pm$0.09 & 0.81$\pm$0.09\\
M/L$_{bulge,I}$ & 1.90$\pm$0.09 & 1.83$\pm$0.09\\
e$_{halo}$ & N/A & 0.01\\
Scale & $r_c$ = 1.03$\pm$0.02kpc & $r_h$ = 12.0$\pm$0.6kpc\\
$\chi^2$ & 1.4 & 1.0\\
\end{tabular}\caption{Best fitting parameter values for the two halo density profiles. Note the similarity between mass-to-light ratio values and the almost spherical NFW halo.}
\end{center}
\end{table}
It is evident from the similarity of the mass-to-light ratios and the small NFW flattening that these two halo profiles would contribute comparable convergence and shear to the lensing. Figure 1 shows the rotation curve of a galaxy with the best-fitting parameters for the softened isothermal sphere halo and luminous components.
\begin{figure}[h]
\plotone{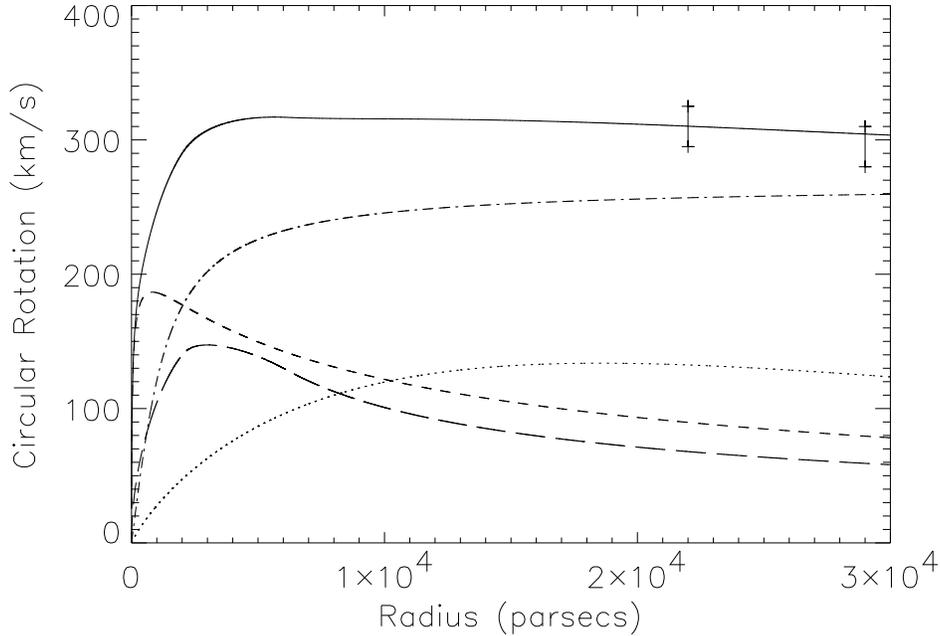}
\caption{Best-fitting rotation curve based upon rotation points from HI observations and image positions for a softened isothermal sphere dark matter halo. The solid line is the total rotation curve, the dash-dotted line is the dark matter halo, the dashed line is the bulge, the dotted is the disk and the long-dashed is the bar.}
\end{figure}
The bulge component clearly dominates the centre of the galaxy, providing the majority of the dynamical support within the image region. In the outer regions of the galaxy the dark matter component dominates, as expected from results of previous work deconvolving galactic rotation curves. The best-fitting rotation curve using the flattened NFW dark matter halo is essentially the same.

The disk is clearly sub-maximal according to the definition of Sackett (1997), however the importance of the bulge in this system should not be forgotten. This early-type spiral is not the ideal system for studying the importance of disk stellar support as the bulge clearly dominates the disk.
\vspace{-4mm}
\section{Discussion}

Although the results presented are acceptable fits to the data, they are not likely to describe the real situation in the galaxy. Firstly, the flux ratios of the images have not been included in the minimisation procedure, even though they are additional constraints. The first reason for this is the uncertain nature of ratios measured for most lensing systems. Q2237+0305 is not a radio-loud quasar and radio measurements are difficult. Optical fluxes probe a small region in the source plane and are subject to possible microlensing leading to spurious values. Agol, Jones \& Blaes (2000) measured mid-IR fluxes for this system arguing the larger extent of the emitting region would reduce the contamination of microlensing and the waveband would minimise the effects of dust reddening. We will use these measurements to compare with our best-fitting model. The second reason for omitting the flux ratios in the analysis is the contention over their applicability when dark matter substructure is taken into account (Schechter \& Wambsganss, this meeting). Our dark matter is modelled as a smooth component.

Comparison of the flux ratios shows excess flux in one of the images with both profiles for the dark matter. This result is quite robust and suggests an underlying problem in the mass distribution.

In addition to this problem, the bulge mass-to-light ratios are extremely low compared with other studies of the bulge component in I-band (see for example Fukugita, Hogan \& Peebles 1998) which place the figure closer to M/L=5.0$\Upsilon$. Clearly, increasing the bulge mass changes the convergence and the shear in the image region due to its dominance of the mass there. Without changing the remainder of the mass components, the image positions would be torqued out of place.

The location of the images defines the projected mass within that region quite accurately (Wambsganss \& Paczynski 1994 measured M$_{enc}$ = (1.48$\pm$0.01) $\times$ 10$^{10}M_{\odot}$h$_{75}^{-1}$) and therefore the bulge mass-to-light ratio can only be increased so far before it alone provides all of the convergence. In addition, as the bulge is an elliptical component, increasing its contribution will increase the shear.

The one completely unknown component of the mass is the dark matter halo. The best-fitting disk and bar sizes are what one would expect. If the bulge is too small in our models then it would appear the dark matter distribution would need to be changed to accommodate this. The most likely outcome is a halo with little contribution within the image region but sufficiently flattened as to counteract the increased torque from the bulge. Without further information, this is purely speculation.

Additional dynamical information can help to answer these remaining questions. With only rotation data for the outer galaxy, critial regions of the rotation curve are unknown. Well measured points in the galactic centre will tightly constrain the allowed dynamics in the important image region, as shown in Figure 2.
\begin{figure}[h]
\plotone{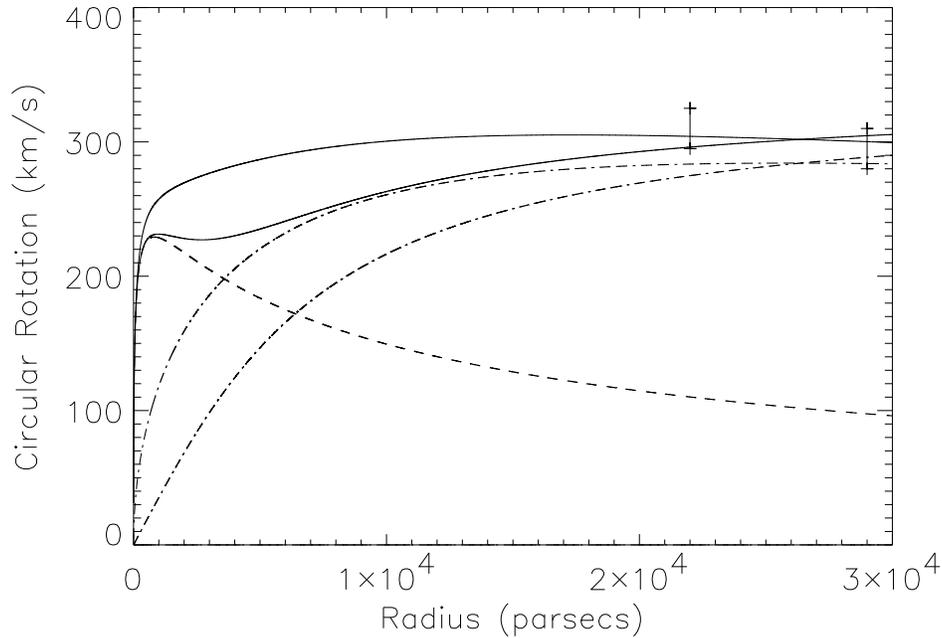}
\caption{Comparison of expected rotation curves for a maximal bulge (M/L$_I$=3.0) with the two different halo profiles. The lower dash-dotted line shows the contribution of the SIS with $r_c$=5kpc and the upper, a spherical NFW with $r_s$=12kpc. The dashed line shows the bulge contribution. The two solid lines are clearly different in the inner regions.}
\end{figure}
It compares the expected observed rotation curve for this galaxy with a maximal bulge and the two different halo profiles. This solution does not fit the image positions, but shows the possible difference in the inner rotation curve. It is this splitting of curves that we hope to probe with future spectroscopy of the galaxy.
\vspace{-6mm}
\section{Conclusions}
We have modelled the galaxy 2237+0305 with multiple mass components to investigate the mass-to-light ratios of the luminous components, the dynamical importance of the disk and bulge and the shape of the dark matter halo. We require additional dynamical information to define the dark matter contribution accurately. This system is potentially useful for constraining the mass-to-light ratio of the bulge component and the shape and inner slope of the dark matter halo.

\end{document}